# Improvement of image-type very-low-energy-electron-diffraction spin polarimeter

Heming Zha,[1,2,*] Wenjing Liu,[1,2,*] Deyang Wang,[1,2,*] Bo Zhao,[4] XiaoPing Shen,[5] Mao Ye,[1,2] and Shan Qiao[1,2,3‡]

*1 Center for Excellence in Superconducting Electronics, State Key Laboratory of Functional Materials for Informatics, Shanghai Institute of Microsystem and Information Technology, Chinese Academy of Sciences, Shanghai 200050, People's Republic of China*
*2 Center of Materials Science and Optoelectronics Engineering, University of Chinese Academy of Sciences, Beijing 100049, People's Republic of China*
*3 School of Physical Science and Technology, ShanghaiTech University, Shanghai 201210, People's Republic of China*
*4 Shanghai Advanced Research Institute, Chinese Academy of Sciences, Shanghai 201210, China*
*5 State Key Laboratory of Surface Physics, Department of Physics, Fudan University, Shanghai 200433, People's Republic of China*

[*]These authors contributed equally to this work.
[‡]The author to whom correspondence may be addressed: qiaoshao@mail.sim.ac.cn

Spin- and angle-resolved photoemission spectroscopy (SARPES) with high efficiency and resolution plays a crucial role in exploring the fine spin-resolved band structures of quantum materials. Here we report the performance of SARPES instrument with a second-generation home-made multichannel very-low-energy-electron-diffraction (VLEED) spin polarimeter. Its energy and angular resolutions achieve 7.2 meV and 0.52°. We present the results of SARPES measurements of Bi(111) film to demonstrate its performance. Combined with the density functional theory (DFT) calculations, the spin polarization of the bulk states was confirmed from the spin-layer locking caused by the local inversion asymmetry. The surface states at binding energy of 0.77 eV are found with $1.0 \pm 0.11$ spin polarization. The better resolutions and stability compared with the first-generation one provide a good platform to investigate the spin-polarized electronic states in materials.

## I. Introduction

Angle-resolved photoelectron spectroscopy, which can measure the energy and momentum of electrons in materials, is one of the powerful tools to investigate the origin of physical phenomena by studying their electronic states. In recent years, high-temperature superconductors,[1,2] giant magnetoresistance effect,[3] and topological materials[4] have been the hotspots of research on condensed matter physics. The peculiar characters of these materials are usually related to spin-orbital or exchange interactions, in which the electron spin plays an integral role. SARPES is the best method to investigate the spin-dependent band structures to explore the mechanism of those peculiar characters.

With the recent development of high-performance electron analyzers, the band structures can be investigated more precisely to give hallmarked specifics such as gaps or kinks to study some problems in physics. For spin-resolved band structure measurements, However, the resolutions and efficiency are still too low to study the fine structures. The Mott-type spin polarimeter achieved an efficiency of only $10^{-4}$.[5,6] The first step to improve the efficiency was the VLEED-type spin polarimeter,[7] whose efficiency was about 100 times higher than that of the Mott-type. The second step was the image-type multichannel spin polarimeter.[8,9] The first-generation image-type VLEED spin polarimeter based on exchange-scattering was developed by our group five years ago. It makes the observations of complex spin structures possible.[10] However, it has many defects that need to be improved.

## II. The design of the second-generation image-type VLEED spin polarimeter

The analyzer in the system is a VG Scienta R3000 hemispherical energy analyzer (HEA) and its specifications of energy and angular resolutions are 3.0 meV and 0.1° with entrance slit width of 0.2 mm and pass energy of 2 eV, respectively. The primary function of the electron optics of the second-generation image-type VLEED polarimeter is the same as that of the first-generation. On the exit plane of the HEA, the photoelectrons with the same energy and emergence angle are focused on the same position, where the positions with different x- and y-coordinates denoted in Fig. 1 correspond to the different energies and emergence angles, and their intensities form an image. As shown in Fig. 1(e), the image is point-to-point transferred by the lenses (LS) 1 and 2 to the ferromagnetic Fe(001)-P(1 × 1)-O target, and the LS 2 and 3 point-to-point transfer the image on the target to the entrance plane of electron detector consisted of two microchannel plates (MCP) and a fluorescence screen. Meanwhile, the electron beam is turned 180 degrees two times by the magnetic field to ensure the normal incidences[11] to the target and electron detector. Under point-to-point transfer, the information of momentum, i.e., emergence angle, and energy of photoelectrons will not be lost. The spin polarization of the electron beam can be obtained by the asymmetry of scattering rates resulting from the exchange interaction for two measurements with opposite magnetizations of the target through a Helmholtz coil. For spin-resolved measurements, the voltage of the target is set as 5.78 V to achieve the maximum efficiency. For spin-integrated measurements, the voltage of the target is adjusted to -1 V to prevent electrons from hitting the target.

In the first-generation polarimeter, the dipole magnetic field was generated by electromagnets shown by the black parts in Fig. 1(c). The different optical characters of dipole magnetic field along two directions, the bending of electron beam in the plane perpendicular to magnetic field and the linear motion along the y direction, make the electron beam cannot be focused to the same point in both x and y directions simultaneously. To compensate for this asymmetry, a couple of correction coils, as indicated by the gray slash parts in Fig. 1(c), were added. The dipole magnetic field was found unstable because of a large current of 18 A needed which results in a rise temperature of the electromagnets and caused an image shift in the x-direction. As illustrated in Fig. 1(c), the electromagnets are situated outside the vacuum. To make the vacuum chamber not distort the magnetic field, only non-magnetic material can be selected. Here stainless steel was chosen instead of mu-metal. However, this arrangement cannot shield the environment magnetic field, which can also introduce instability in electron optics. Furthermore, the gap $d_1$ as shown in Fig. 1(c) between the electromagnets is as large as 60 mm, resulting in a strong edge effect that the spatial variation of the magnetic field at the edge area can cause the electrons, initially moving horizontally in x-z plane, to experience a vertical Lorentz force along y direction, which results in the increase of aberrations. Moreover, the magnetic field of magnets with large gap can extend into the nearby electric lenses, causing the deflection of electrons in the lenses to deviate from their axes, which also can result in the increase of aberrations. In addition, after the construction, we found that deflectors were needed to compensate for the alignment errors. However, because of the space limitation, the deflectors could only be set in the magnetic field as shown by the red parts in Fig. 1(c).

The second-generation polarimeter was constructed to overcome the above insufficiencies. Firstly, the electromagnets were replaced with permanent ones to make the magnetic field more stable. The whole polarimeter was enclosed by mu-metal vacuum chamber to shield the environment magnetic field. Meanwhile, the magnet poles were extended into the vacuum chamber to reduce the gap to attenuate the edge effect and the magnetic field inside the nearby lenses. Some lens elements were divided into four parts as shown by the oblique gray lines and the inset in Fig. 1(e). Its functions are twofold, acting as a quadrupole lens to compensate for the different optical characters of dipole magnet along x and y directions and as deflectors to compensate for the alignment errors.

### III. The performance characterization

The size of the entrance aperture of spin polarimeter on exit plane of the HEA is $20 \times 20$ mm$^2$, corresponding to an acceptance angle of 20° and an energy window of 150 meV for 2 eV pass energy. To examine the energy resolution of the spectrometer, the SARPES measurement of Fermi edge of polycrystalline Au was observed as shown in Fig. 2(a) with an analyzer pass energy of 2 eV and an entrance slit width of 0.2 mm. The sample was Au column bombarded by Ar ions *in situ* and its temperature was kept at 5.6 K during the measurements. The instrumental energy resolution was determined by fitting the Fermi edge as shown in Fig. 2(c). The fitting process involved the convolution of a Gaussian function with the addition of a linear background and the Fermi-Dirac function at the sample temperature. The full width at half maximum (FWHM) of the Gaussian function provided a direct measurement of the instrumental resolution. Based on this fitting procedure, the instrumental resolution was found to be 7.2 ± 0.12 meV. To evaluate the angular resolution, an aperture with sharp edge was set between the sample of a 40 μm diameter Au wire and the analyzer and the distance between the sample and the aperture is 7 mm. The SARPES spectrum for different emergence angles was observed as shown in Fig. 2(b). In general, the angular resolution refers to the ability of the system to distinguish between two beams with close emission angles. After a knife edge aperture, the ideal case of infinite angular resolution corresponds to an intensity distribution of a step function along the angular direction. In this ideal scenario, the corresponding first-order derivative of the intensity distribution is a delta function. However, in practical situation, the finite resolution of the system leads to a broadening of the intensity distribution.

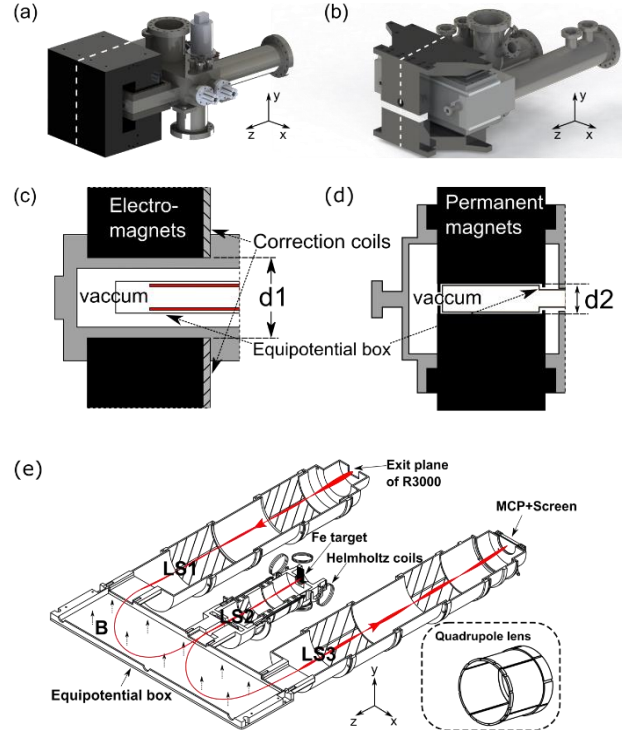

FIG. 1. Diagram of VLEED-type polarimeter structures. 3D schematic diagram of the (a) first-generation and (b) second-generation polarimeters, (c), (d) The cross-section along the white dash lines shown in (a) and (b) respectively and the black parts present the magnets. In (c), the gray slash part is the correction coils and the red part is the deflection electrodes. (e) Schematic drawing of second-generation VLEED polarimeter. The equipotential box is set to ensure that the electrons have a certain speed and go through a perfect circular orbit with a suitable radius. The bunch of red lines shows the schematic electron trajectories. The slash parts refer to the quadrupole lenses. Inset shows the schematic of the quadrupole lens. The Fe target was magnetized by Helmholtz coil in x or y direction.

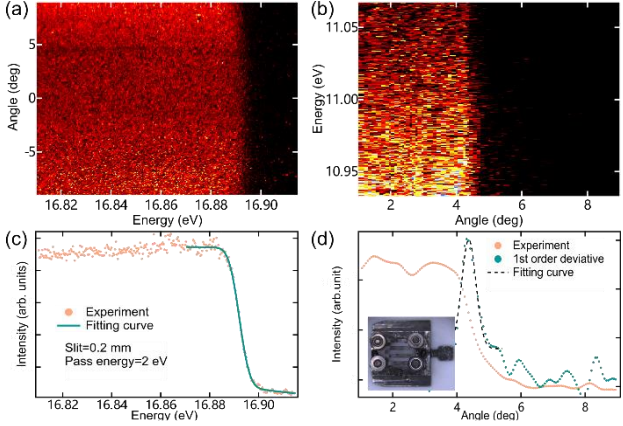

FIG. 2. (a), (b) E-k image of polycrystalline Au and Angular device respectively. (c) Fermi distribution curve measured in fix mode with polycrystalline Au. (d) the intensity distribution cut from (b) along the angle direction. Inset shows the schematic of the angular device.

Consequently, the intensity distribution can no longer be described as a step function. In this case, the first-order derivative of the broadened intensity distribution can be fitted using a Gaussian function. The FWHM of this Gaussian function can then be used as the measurement of the angular resolution of the system. Here, by fitting the first-order derivative curve by Gaussian function, as shown in Fig. 2(d), the angular resolution of the SARPES instrument was evaluated to be about 0.52°.

## IV. SARPES measurements of Bi(111)

Bi(111) film grown on Si(111) substrate is a typical material with Rashba-splitting surface states, which is suitable to demonstrate and quantitatively character the performance of the polarimeter. We prepared the film by molecular beam epitaxy and transfer it to the analysis chamber *in situ*[12] with an ultrahigh vacuum of $10^{-8}$ Pa. The pass energy of HEA was set as 2 eV and the slit of HEA was set to 0.8 mm in following experiments for more photoelectron intensity corresponding to a resolution of 12 meV in spin-resolved mode.

Fig. 3(a) shows the spin-integrated Fermi surface mapping of Bi(111) with 20 meV integration window and 0.25° angular step. There are six petal-shaped electron pockets along $\bar{\Gamma} - \bar{M}$ directions and an electron pocket at $\bar{\Gamma}$ point. The energy-momentum dispersion along $\bar{M} - \bar{\Gamma} - \bar{M}$ (Fig. 3b) shows the Rashba-type surface state $\alpha$ near the Fermi surface and another surface state $\gamma$ at binding energy of 0.5 eV to 0.8 eV which are consistent with the previous reports.[13,14] Fig. 3(c) and (d) compare the spin-resolved E-k images along $\bar{\Gamma} - \bar{M}$ direction with the spin direction in plane and perpendicular to momentum measured by the first- and second-generation VLEED polarimeters, respectively. Through the comparison of the images, we can intuitively find the significant improvement of second-generation one in resolutions with clearer spin splitting. The bulk state $\beta$ at the binding energy of 0.25 eV also shows a clear polarization that we did not observe in the first-generation one. The effective Sherman function ($S_{eff}$) of the system can be obtained from $S_{eff} = A/P$, where P presents the spin polarization of injected electrons and asymmetry $A = (I_+ - I_-)/(I_+ + I_-)$ is measured from the intensities $I_+$ and $I_-$ of reflected electrons with reversed magnetizations of iron target. The momentum distribution curves of asymmetries of $\alpha$ and $\gamma$ bands are shown in Figs. 3(e) and (f) along the cuts indicated by thick dashed lines A and B shown in Fig. 3(d), and the maximum asymmetries of $\alpha$ and $\gamma$ bands were found to be 0.33 ± 0.023 and 0.35 ± 0.036 respectively. The effective Sherman function can be determined by assuming the 100 % polarization of $\alpha$ band,[15] and the effective Sherman function of the spin polarimeter can be determined as 0.33. The efficiency of the spin polarimeter is $2.2 \times 10^{-2}$, which is termed as figure of merit (FOM) $\varepsilon = S_{eff}^2 I/I_0$, where $I$ and $I_0$ are intensities of scattered and incident electrons and $I/I_0$ was measured as 0.20. From the results, the polarization of the $\gamma$ band in Fig. 3(f) can be determined as 1.0 ± 0.11 at -0.082 Å$^{-1}$ and -1.1 ± 0.11 at 0.083 Å$^{-1}$. The spin-orbital coupling constant $\alpha_R$ (Rashbar parameter) was determined by linear fitting of the splitting magnitude to be 1.69 eVÅ$^{-1}$ for the $\gamma$ band, which is about 3 times that of the $\alpha$ band.[9] With both 100 % spin polarization and larger splitting, it is more convenient to choose the $\gamma$ band at binding energy of 0.77 eV to evaluate the effective Sherman function in the future to avoid the underestimation of $S_{eff}$ caused by the poor energy resolution. Need to be mentioned, the acquisition time of SARPES image shown in Fig. 3(d) is only two hours in total.

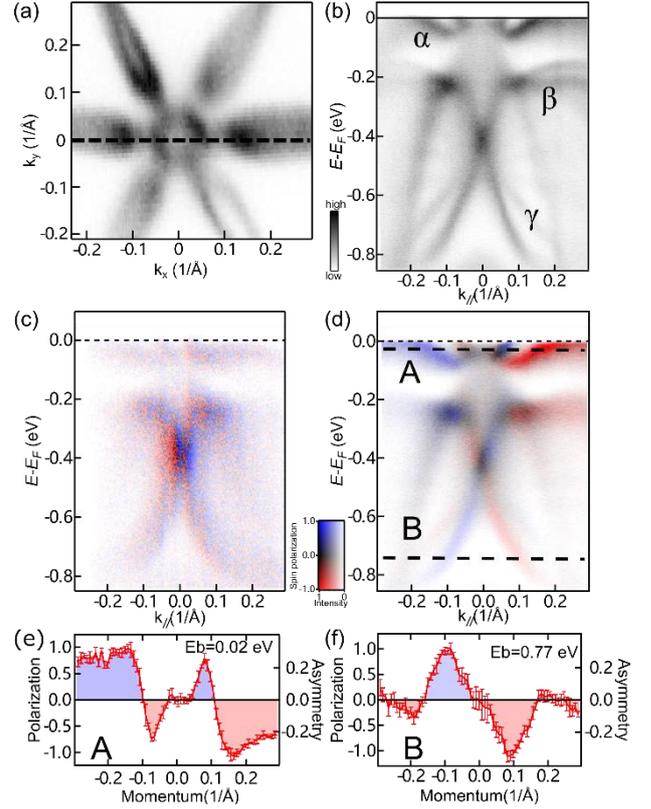

FIG. 3. (a) Spin-integrated Fermi surface map near $\bar{\Gamma}$ point at 6.0 K. (b) Spin-integrated band dispersion along cut marked by the dashed line in figure (a). (c), (d) Spin-resolved E-k images with the spin direction in-plane and perpendicular to momentum measured by first- and second-generation VLEED polarimeters respectively. (e), (f)The momentum distribution curves at binding energy of 0.02 eV and 0.77 eV obtained from figure (d) and the cuts of binding energies are indicated by the thick dashed lines. The momentum distribution curve here is obtained by integration of spectra in 8 meV and 0.23 degrees ranges in the energy and angular directions. Error bars are standard deviations of the results from five consecutive measurements.

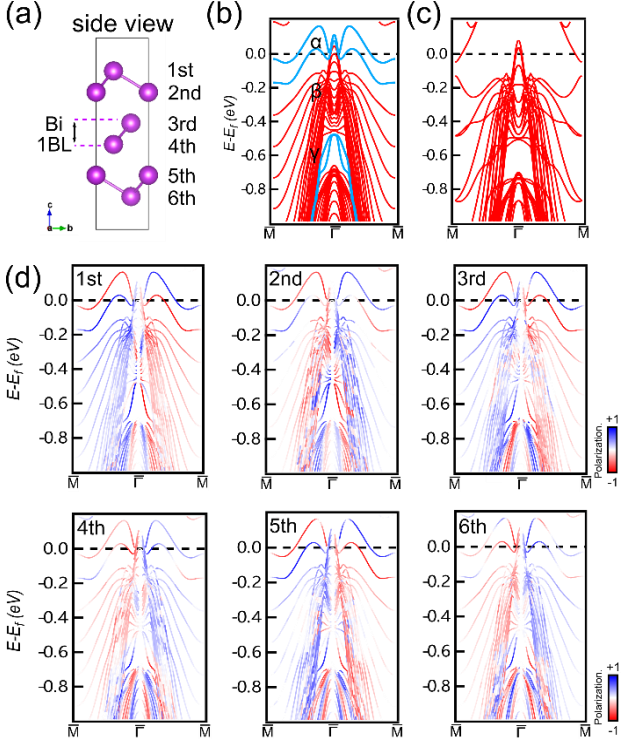

FIG. 4. The calculations of the band dispersion of 18-bilayer slab model. (a) Side view of atomic structure of Bi thin film. (b) Results with the vacuum layer in which the surface components are denoted by azure color curves. (c) Results without the vacuum layer. (d) The spin polarizations of top six atomic layers along $\bar{M}-\bar{\Gamma}-\bar{M}$ direction extracted from slab calculations.

First-principle calculations were done to confirm the origin of the spin polarization and verify that the $\alpha$ and $\gamma$ bands have 100 % polarizations. The DFT calculations are performed by the Vienna ab initio simulation package (VASP) with the Perdew-Burke-Ernzerhof (PBE) method.[16-18] A kinetic energy cutoff of 500 eV and a uniform $9 \times 9 \times 1$ k-point grid in the Brillouin zone were proved to be sufficiently accurate for the calculations of the bands. Fig. 4(a) shows the side view of the slab model of 3 bilayer (BL) Bi, from the top first atom to the bottom sixth atom. Fig. 4(b) and (c) are the calculation results of the 18 BL slab model with and without the vacuum layer. The azure curves in Fig. 4(b) can be determined as surface states corresponding to the observed $\alpha$ and $\gamma$ bands shown in Fig. 3(d). As shown in Fig. 4(d), in-plane spin polarizations perpendicular to the momentum of bulk states of the first six Bi atomic layers are layer dependent. Though the crystal structure of Bi is centrosymmetric, the layer dependent spin polarizations confirm that the spin polarizations of bulk states result from the local inversion asymmetry with the Bi-site point group ($C_{3v}$). The polarization of spin-layer locked state can be compensated by its inversion counterpart with opposite polarization to ensure the whole energy band remains doubly Kramer degeneracy.[10,19-21] Under the photon energy of 21.2 eV, the detection depth of photoelectrons is about 0.6 nm which corresponds to nearly 1.5 BL. Thus, we can observe the residual spin polarization of $\beta$ bands attributed to the topmost three atoms here. From the above discussions, it is clear that the experimental and calculated results are consistent with each other.

## V. SUMMARY

The first-generation image-type VLEED spin polarimeter is upgraded by using permanent magnets and quadrupole lenses. The effective Sherman function was determined by measuring Bi(111)/Si(111). Combined with the density functional theory calculations, we confirmed that the spin polarization of bulk $\beta$ band originates from the spin-layer locking caused by the local inversion asymmetry and both the $\alpha$ and $\gamma$ surface states have 100 % spin polarizations. The excellent performance and stability of second-generation VLEED spin polarimeter provide a platform to investigate the complex spin-polarized electronic states in materials.

## ACKNOWLEDGMENTS

This work is supported by the National Key R&D Program of China (2022YFB3608000), and National Natural Science Foundation of China (No. U1632266, No. 11927807, No. U2032207).

## DATA AVAILABILITY STATEMENT

The data that support the findings of this study are available from the corresponding author upon reasonable request.